\documentclass[aps,prb,twocolumn,floatfix]{revtex4}
\usepackage{amsfonts}
\usepackage{amsmath}
\usepackage{amssymb}
\usepackage{graphicx}

\input{tcilatex}

\begin{document}

\title{Magnetic irreversibility and Verwey transition in nano-crystalline
bacterial magnetite}
\author{Ruslan Prozorov}
\affiliation{Ames Laboratory and Department of Physics \& Astronomy, Iowa State
University, Ames IA 50011, USA}
\author{Tanya Prozorov}
\affiliation{Department of Chemical and Biological Engineering, Iowa State University and
Ames Laboratory, Ames IA 50011}
\author{Timothy J. Williams}
\affiliation{Department of Biochemistry, Biophysics and Molecular Biology, Iowa State
University, Ames IA 50011}
\author{Dennis A. Bazylinski}
\affiliation{School of Life Sciences, University of Nevada, Las Vegas, 4505 Maryland
Parkway, Las Vegas, NV 89154-4004}
\author{Surya K. Mallapragada}
\affiliation{Department of Chemical and Biological Engineering, Iowa State University and
Ames Laboratory, Ames IA 50011}
\author{Balaji Narasimhan}
\affiliation{Department of Chemical and Biological Engineering, Iowa State University and
Ames Laboratory, Ames IA 50011}
\date{18 May 2007}

\begin{abstract}
The magnetic properties of biologically-produced magnetite nanocrystals
biomineralized by four different magnetotactic bacteria were compared to
those of synthetic magnetite nanocrystals and large, high quality single
crystals. The magnetic feature at the Verwey temperature, $T_{V}$, was
clearly seen in all nanocrystals, although its sharpness depended on the
shape of individual nanoparticles and whether or not the particles were
arranged in magnetosome chains. The transition was broader in the individual
superparamagnetic nanoparticles for which $T_{B}<T_{V}$, where $T_{B}$ is
the superparamagnetic blocking temperature. For nanocrystals organized in
chains, the effective blocking temperature $T_{B}>T_{V}$ and the Verwey
transition is sharply defined. No correlation between particle size and $%
T_{V}$ was found. Furthermore, measurements of $M\left(H,T,time\right)$
suggest that magnetosome chains behave as long magnetic dipoles where the
local magnetic field is directed along the chain. This result confirms that
time-logarithmic magnetic relaxation is due to the collective (dipolar)
nature of the barrier for magnetic moment reorientation.
\end{abstract}

\maketitle

\section{Introduction}

Magnetite is one of the most studied ferrimagnetic compounds. A sudden
change in its thermodynamic properties above $100$ K has been the point of
interest for many years. In 1926, Parks and Kelley reported a significant
heat absorption in magnetite at about $113-117$ K \cite{parks}. They
speculated that this is due to a change in the magnetic subsystem and not in
the crystal structure. Thirteen years later, Verwey found that resistivity
in magnetite increases about two orders of magnitude upon cooling below $120$
K, which he attributed to the order-disorder transition in the electronic
subsystem \cite{verwey39}. Initially confirmed, later this model was
questioned and massive experimental and theoretical effort was mounted to
investigate the problem. The literature on the Verwey transition is vast and
we refer the reader to several reviews \cite{review,muxworthy,garcia,leonov}.

Magnetite has an inverse spinel crystal structure, Fe$_3$O$_4$=$\left[\text{%
Fe}^{3+}\right] _{A}\left[ \text{Fe}^{3+}\text{Fe}^{2+}\right]_{B}$O$_{4}$,
where $A$ sites are coordinated in tetrahedra and $B$ sites are coordinated
in octahedra. Magnetic order at the $A$ and $B$ sites is antiparallel
resulting in ferrimagnetism with an excess magnetic moment of about $4\mu_{B}
$ per formula. (For each formula unit there are two $B$ sites with spin $%
S=2.25$ and one $A$ site with $S=2.5$). There are eight formula units in the
cubic cell (with cell constant 8.4 \AA ), so each cubic unit cell
contributes $32\mu_{B}$. Above the Verwey temperature, $T_{V}\simeq120$ K,
the $B$ sites are charge - frustrated so electrons are significantly
delocalized, which leads to moderate conductivity of about $0.01$ $%
\Omega\cdot$cm at room temperature. Below $T_{V}$, resistivity increases two
orders of magnitude, which Verwey explained in terms of charge ordering in
the $B$ subsystem: Fe$^{2+}$ ions along $\left[ 110\right] $ and Fe$^{3+}$
along $\left[ 1\overline {1}0\right] $ directions \cite{verwey41} (the
"Verwey model").

In this paper, we discuss the magnetic signature of this transition that
appears as a very sharp feature (in good single crystals \cite{wang90,kakol}%
) at the Verwey temperature, $T_{V}$. Similar to the Verwey transition
itself, the physics of this magnetic anomaly is still debatable. It is known
that magnetic easy axis changes from the $\left\langle 111\right\rangle $ to 
$\left\langle 100\right\rangle $ direction (within 0.2$^{o}$ due to
monoclinic distortion) below $T_{V}$ \cite{lima}. $K_{1}$ anisotropy energy
increases by an order of magnitude, but the amplitude of the magnetic moment
does not change (see Fig.~\ref{fig3}).

In recent years, much attention was drawn to nanoparticles of magnetite in
an attempt to understand the Verwey transition and the behavior of
ferromagnetic nanoparticles. Various, often contradictory results have been
reported. Disappearance of the Verwey transition was suggested to occur in
nanoparticles grown under a weak (0.25 T) magnetic field \cite{wang}.
Significant reduction of the Verwey transition temperature, $T_{V}$, was
reported in relatively large nanoparticles ($T_{V}\simeq20$ K in 50 nm
nanoparticles) \cite{goya}. No transition was observed in 7-10 nm
nanoparticles and it was even suggested that monoclinic distortion does not
develop in such particles \cite{goya,arelaro}. On the other hand, direct
transport measurements suggest that the transition in resistivity exists
above $100$ K both in stacked and individual nanoparticles with significant
magneto-resistance peak above the gap in the $I-V$ curves \cite{markovich}.
Overall, the matter is complicated by various interpretations of the
superparamagnetic regime, superparamagnetic blocking, magnetic relaxation
and the difference between single-particle and collective behavior in
ferromagnetic nanoparticles. In addition, magnetite, especially in form of
nanoparticles, is very sensitive to oxidation by oxygen that might result in
stoichiometric changes in the crystals and disturbance of the transition 
\cite{ozdemir}.

In this paper, we compare biologically-produced magnetite nanocrystals,
which have an almost perfect crystal structure and a well-defined shape and
size, to large high-quality crystals \cite{wang90,kakol} as well as to
conventional, synthetic magnetite nanoparticles. The goal was to study the
Verwey transition in a superparamagnetic system (without magnetic domains)
and identify possible influences of particle size, shape and crystalline
magnetic anisotropies as well as interparticle interactions on the magnetic
signature of the Verwey transition. An additional influencing factor is that
the bacterial magnetite crystals are enveloped by a phospholipid membrane
that protects the particles from oxidation.

Magnetotaxis in bacteria was first reported in 1975 \cite{blakemore} and
magnetite crystals were identified in these bacteria in 1979 \cite{frankel75}%
. Since then, various properties of different magnetotactic bacteria have
been extensively studied \cite%
{krueger,moskowitz,meldrum,penninga,bazylinski,frankel,weiss,faivre,pan,polyakova,simpson,kobayashi,weyland,bazylinski07}%
. Most of the studies were performed on the bacteria extracted from natural
aquatic habitats. Recently significant progress was achieved \textit{in vitro%
} studying the synthesis of magnetite by various bacterial strains \cite%
{bazylinski}. This allowed for the targeted modification of magnetic
properties and various post-synthesis modifications. For reviews of
magnetite formation in prokaryotes as well as on the general ecophysiology
of magnetotactic bacteria, see \cite{bazylinski,bazylinski07}. While the
majority of prior studies have focused on various aspect of magnetic
behavior of different bacterial nanoparticles in terms of the influence of
dipolar interactions, size and shape effects, size distribution, orientation
of magnetic moments with respect to the magnetosome chains and particles
themselves etc., here we use bacterial magnetite to understand how the
magnetic signature of the Verwey transition reveals itself in nanoparticles
of various sizes and shapes as well as in those self-assembled in
magnetosome chains and compare it to that of bulk single crystals. Off-axis
electron holography suggests an almost perfect alignment of local magnetic
induction along the magnetosome chain above the Verwey temperature, $T_{V}$ 
\cite{simpson}. This direction is also the $\left[ 111\right] $
crystallographical direction of each nanocrystal in the chain. Below $T_{V}$%
, the magnetic induction develops some undulation due to change of the
magnetic easy axis to the $\left\langle 001\right\rangle $ direction.
However, there is still magnetic coherence between the individual particles
in the chain. Our results show that the Verwey temperature is not
significantly affected in nanoparticles, but its signature is very sensitive
to the magnetocrystalline and shape anisotropies that determine the magnetic
blocking temperature.

\begin{figure*}[t]
\begin{center}
\includegraphics[width=15cm]{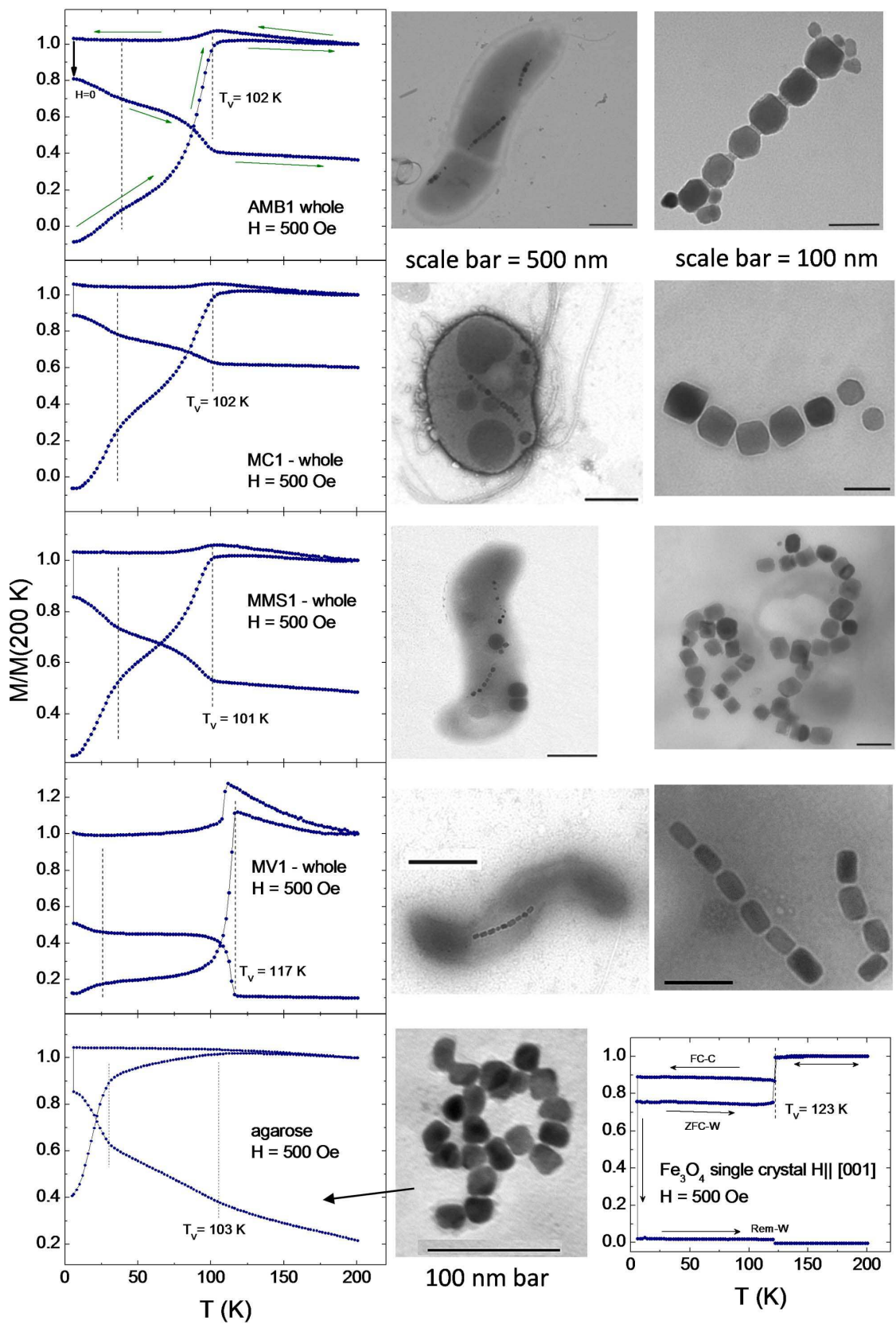}
\end{center}
\caption{$M\left(T\right)$ measurements of various forms of Fe$_{3}$O$_{4}$
nanoparticles. In each case, three measurements, shown here as three
separate curves, were taken as described in the text - ZFC-W and FC-C at $%
H=500$ Oe, and then turning the field off and warming of the remanent
magnetization. The top four rows correspond to biologically-produced
magnetite from four different whole bacterial cells - strains AMB-1, MC-1,
MMS-1 and MV-1, respectively. For comparison, the lower left frame shows
measurements of synthetic Fe$_{3}$O$_{4}$ nanoparticles and the lower right
frame shows data from large single crystals. The middle column, except for
the bottom figure, are transmission electron micrographs (TEMs) of whole
cells of the magnetotactic bacterial strains. The bottom image is a TEM of
synthetic magnetite. The right column, except for the bottom figure, are
TEMs of magnetosome chains released from lysed cells.}
\label{fig1}
\end{figure*}

\section{Experimental}

\subsection{Growth of magnetotactic bacterial strains}

Four magnetotactic bacterial strains were investigated: strain MV-1
(referred to as wild-type (WT) MV-1), cells of which are vibrioid
(curved)-to-helical in shape \cite{bazylinski88}; strain MC-1, a marine
coccus (roughly spherical) \cite{williams06}; strain MMS-1, with cells that
range in shape from vibrioid-to-helical \cite{bazylinski88,williams06}; and 
\textit{Magnetospirillum magneticum} strain AMB-1, a spirillum that is
generally always helical \cite{matsunaga}. Also examined was a
non-magnetotactic mutant strain of MV-1, called MV-1nm-1 \cite{dubbels},
identical in cellular morphology to (WT) MV-1 but devoid of intracellular
magnetite chains.

\subsubsection{Strain MV-1 (including mutant)}

Cells of strain MV-1 were grown anaerobically in 1.2 liter of liquid media
in 2 liter glass bottles. The medium consisted of an artificial sea water
(ASW) base \cite{bazylinski} to which was added (per liter) prior to
autoclaving: 0.2 ml 0.2$\%$ aqueous resazurin; 5.0 ml modified Wolfe's
mineral elixir \cite{frankel,wolin}; 0.5 g sodium succinate$\times$6H$_{2}$%
O; 0.2 g sodium acetate$\times$3H$_{2}$O; 0.5 g CasAmino Acids (\textit{%
Difco Laboratories}, Detroit, Mich., USA); 0.25 g NH$_{4}$Cl; and 100 $\mu$l
0.2$\%$ (w/v) aqueous resazurin. The pH of the medium was adjusted to 7.0.
Bottles were sealed, and the medium then bubbled with N$_{2}$ gas for an
hour, followed by N$_{2}$O gas for one hour. (Flow rate of all gases and gas
mixtures was approximately 100 ml/min.) The medium was then autoclaved.
After autoclaving and cooling, the following solutions were injected into
the medium bottles from anaerobic stocks (except for cysteine, which was
made fresh and filter sterilized directly into the medium), in order: 1.8 ml
of 0.5 M KHPO$_{4}$ buffer, pH 7.0; 2.0 ml of neutralized 0.43 M cysteine$%
\times $HCl$\times$H$_{2}$O; and 2.9 ml of 0.8 M NaHCO$_{3}$. The medium was
allowed to chemically reduce (become colorless) after which 2.9 ml of 0.01 M
FeSO$_{4}$ (dissolved in 0.02 M HCl) and 0.6 ml of vitamin solution \cite%
{frankel,wolin} was added. The medium was inoculated and then incubated at
28 $^o$C for approximately one week at which time cultures had reached the
end of exponential growth.

\subsubsection{Strains MC-1 and MMS-1}

Strains MC-1 and MMS-1 were grown separately in identical liquid media.
Cells were grown microaerobically in 850 ml of media in 2 liter glass
bottles. The medium consisted of the same ASW base as described above, to
which was added (per liter) prior to autoclaving: 5 ml modified Wolfe's
mineral elixir; 0.25 g NH$_{4}$Cl; and 100 $\mu$l 0.2$\%$ (w/v) aqueous
resazurin. The pH of the medium was adjusted to 7.0, and 1.07 g NaHCO$_{3}$
added. The bottles were then sealed and bubbled with 7.5$\%$ CO$_{2}$ gas in
N$_{2}$ (flow rate about 100 ml/min) passed over heated copper wire to
remove O$_{2}$ for one hour. The bottles were sealed and autoclaved. After
autoclaving and cooling, the following solutions were injected into the
media bottles from anaerobic stocks (except for cysteine), in order: 1.5 ml
of 0.5 M KHPO$_{4}$ buffer pH 6.9, 1 ml of 0.23 M neutralized cysteine$\times
$HCl$\times$H$_{2}$O; 10 ml of 25$\%$ (w/v) Na$_{2}$S$_{2}$O$_{3}\times$5H$%
_{2}$O; and 0.4 ml of vitamin solution (as above). The medium was allowed to
reduce after which 2.5 ml of 0.01 M FeSO$_{4}$ (dissolved in 0.02 M HCl) was
injected. The medium was inoculated, after which 6 ml of sterile O$_{2}$ was
injected (0.4$\%$ of the headspace), and the bottles carefully placed
without shaking so as not to disturb the forming O$_{2}$ gradient, at 25 $^o$%
C. As soon as bacterial growth was evident (as a scum on the surface of the
medium), sterile O$_{2}$ was injected into the headspace at regular
intervals to promote further growth of the cells.

\subsubsection{\textit{Magnetospirillum magneticum} strain AMB-1}

Cells of \textit{M. magneticum} strain AMB-1 were grown anaerobically in 1.2
liter of liquid medium in 2 liter glass bottles. The medium consisted of
(per liter): 5 ml modified Wolfe's mineral elixir; 10 ml Wolfe's vitamin
solution \cite{frankel}; 0.68 g KH$_{2}$PO$_{4}$; 0.85 g sodium succinate$%
\times $6H$_{2}$O; 0.58 g sodium tartrate$\times$2H$_{2}$O; 0.083 g sodium
acetate$\times$3H$_{2}$O; 225 $\mu l$ 0.2$\%$ (w/v) aqueous resazurin; 0.17
g NaNO$_{3}$; 0.04 g ascorbic acid; 2 ml of 0.01 M ferric quinate \cite%
{blakemore79} (made by combining 0.19 g quinic acid and 0.27 g of FeCl$%
_{3}\times$6H$_{2}$O in 100 ml distilled, deionized H$_{2}$O). The pH of the
medium was adjusted to 6.75. Bottles were sealed, then the medium sparged
with N$_{2}$ for one hour. The medium was then autoclaved. When cool, the
medium was inoculated and the culture incubated at 28 $^o$C.

\subsubsection{Harvesting, lysis and drying of bacterial magnetite}

Bacterial cells were harvested in the late exponential stage of growth by
centrifuging cultures at 6000 rpm for 15 min at 4 $^o$C. Cell suspensions
were prepared by resuspending the centrifuged cells in ice-cold, sterile ASW
buffered with 10 mM Tris$\times$HCl pH 7.0. Cells were lysed by passing cell
suspensions twice through a French pressure cell at 124 MPa. For further
processing, membranes were removed from magnetosomes by treating them with
the strong surfactant, sodium dodecyl sulfate, and the residual magnetite
powder was dried under N$_{2}$ gas in a glovebox at the room temperature.
Measurements on powders and powders re-suspended in water produced similar
results. Moreover, we remeasured the magnetic properties of whole bacteria,
lysed cells and dried cells left in gelatin capsules in a refrigerator for
more than three months and did not observe changes in any of their magnetic
properties. This served as an indication that particles did not undergo
significant oxidation during the storage period.

\subsection{Magnetite synthesis in agarose gel}

In order to slow the diffusion rates of the reagents to presumably imitate
the conditions under which magnetite nanocrystals are formed in
magnetotactic bacteria, magnetite synthesis was carried out in agarose gels.
All solutions were rendered anaerobic by degassing and sparging with argon
prior to their use. FeCl$_{3}\times6$H$_{2}$O (Aldrich) and FeCl$_{2}\times4$%
H$_{2}$O (Aldrich) were transferred to a reaction flask and dissolved in
water to form a solution with a 1:2 molar ratio of ferrous to ferric ions
(i.e., $0.66$M FeCl$_{3}$ and $0.33$M FeCl$_{2}$ solutions, respectively).
Synthesis of magnetite nanoparticles was carried out via co-precipitation of
FeCl$_{2}$ and FeCl$_{3}$ from aqueous solutions. In a sealed 10 ml
roundbottom flask, agarose (Fisher) was degassed for several minutes and
mixed with one ml of degassed water to prepare a $1\%$ (w/w) solution. The
solution was then heated to boiling under a continuous flow of argon. 100 $%
\mu$l of the 0.66M FeCl$_{3}$ and 0.33M FeCl$_{2}$ solution was added to the
flask and then a drop of 0.0016 M of HCl. The mixture was vigorously stirred
and sparged with argon for 1 minute. The resulting bright yellow solution
was then brought to room temperature, to allow for gelation. After gelation
was completed, 1.5 ml of 0.1 M NaOH was added under under argon. A thin
black band formed at the gel-NaOH interface indicated formation of
magnetite. Nanoparticles of magnetite were allowed to precipitate and
increase in size at room temperature in the sealed flask for 7 days. During
this time period, the thin black band expanded. An aliquot of gel containing
magnetite nanoparticles was taken for magnetic measurements and microscopic
examination. The particles were found to be crystalline and powder x-ray
diffraction confirmed that magnetite was the main crystalline phase.

\subsection{Single crystals of Fe$_{3}$O$_{4}$}

Single crystals of magnetite were synthesized by use of the skull melting
technique and were subsequently annealed to control the oxygen/metal ratio,
as detailed in Ref.\cite{wang90}. A detailed study of the magnetic
properties of these crystals is reported in Ref.\cite{kakol}.

\subsection{Samples and characterization techniques}

The cellular morphology of bacterial strains, magnetosome magnetite particle
size, and the number of magnetosomes per cell or chain were determined by
electron microscopy using a JEOL 1200EX transmission electron microscope
(TEM) at an accelerating voltage of 80 kV. To examine cells, a drop of a
diluted dense bacterial cell suspension was placed on a carbon-coated holey
copper grid. Cell solution was allowed to set for several minutes, after
which time the grid was washed with a drop of water, carefully blotted, and
dried at room temperature. No staining of the grids was performed.

Magnetization measurements were performed using a \textit{Quantum Design}
MPMS magnetometer. For bacterial cells, after centrifugation, a sample of
the bacterial cell suspensions was injected into a waterproof polycarbonate
capsule and immediately cooled below the freezing temperature of the liquid.

Physical characteristics of the samples used in our study are summarized in
Table I. Here $T_{V}$ is the Verwey transition temperature determined by the
high-temperature feature on the $M\left( T\right) $ curves obtained on whole
cells of bacteria as discussed below. Slope represents the slope of the
transition from low- to high-temperature part of the zero-field cooled
curve, $slope$ $=d\left( M/M\left( 200\text{ K}\right) \right) /dT$, which
gives an estimate of the sharpness of the magnetic signature at the Verwey
transition.

\begin{table}[ptb]
\caption{Parameters of studied samples.}
\label{samples}%
\begin{tabular}{c|c|c|c|c}
\hline
sample & volume & dimensions & $T_{V}$ & slope \\ \hline\hline
& (nm$^{3}$) & (nm)$^{3}$ & (K) & K$^{-1}$ \\ \hline\hline
\multicolumn{1}{l|}{AMB-1} & \multicolumn{1}{|l|}{$2\times 10^{5}$} & 
\multicolumn{1}{|l|}{cube, $\left( 55\pm 8\right) ^{3}$} & 
\multicolumn{1}{|l|}{102} & \multicolumn{1}{|l}{0.03} \\ \hline
\multicolumn{1}{l|}{MC-1} & \multicolumn{1}{|l|}{$3\times 10^{5}$} & 
\multicolumn{1}{|l|}{cube, $\left( 70\pm 11\right) ^{3}$} & 
\multicolumn{1}{|l|}{102} & \multicolumn{1}{|l}{0.01} \\ \hline
\multicolumn{1}{l|}{MMS-1} & \multicolumn{1}{|l|}{$2\times 10^{5}$} & 
\multicolumn{1}{|l|}{cube, $\left( 54\pm 9\right) ^{3}$} & 
\multicolumn{1}{|l|}{101} & \multicolumn{1}{|l}{0.01} \\ \hline
\multicolumn{1}{l|}{MV-1} & \multicolumn{1}{|l|}{$1\times 10^{5}$} & 
\multicolumn{1}{|l|}{parallelepiped,} & \multicolumn{1}{|l|}{117} & 
\multicolumn{1}{|l}{0.11} \\ 
\multicolumn{1}{l|}{} & \multicolumn{1}{|l|}{} & \multicolumn{1}{|l|}{$%
\left( 62\pm 8\right) \times \left( 40\pm 6\right) ^{2}$} & 
\multicolumn{1}{|l|}{} & \multicolumn{1}{|l}{} \\ \hline
\multicolumn{1}{l|}{agarose} & \multicolumn{1}{|l|}{$2\times 10^{4}$} & 
\multicolumn{1}{|l|}{cube $\left( 25\pm 3\right) ^{3}$} & 
\multicolumn{1}{|l|}{104} & \multicolumn{1}{|l}{n/a} \\ \hline
\multicolumn{1}{l|}{crystal} & \multicolumn{1}{|l|}{$3\times 10^{17}$} & 
\multicolumn{1}{|l|}{$1.9\times 0.25\times 0.7$mm$^{3}$} & 
\multicolumn{1}{|l|}{125} & \multicolumn{1}{|l}{0.27} \\ \hline
\end{tabular}
\newline
\end{table}

\section{Results}

Various magnetic measurements were performed on ferromagnetic nanoparticles
produced by different systems. In order to compare different samples, we
chose our measurements based on previous studies with various iron oxide
nanoparticles \cite{prozorov99,prozorov04,snezhko05,cao97,cao97a,cao97b} and
report measurements of $M\left( T\right) $ and $M\left( H\right) $
dependencies as well as magnetic relaxation, $M\left(time\right)$, for a
fixed set of parameters as described in detail in the following sections.

\subsection{Temperature dependence of magnetization}

After samples were cooled in a zero magnetic field to $T=5$ K (zero-field
cooling, ZFC), a magnetic field was applied and the temperature dependent
magnetization was measured upon warming (ZFC-W process). Although we tested
a range of field strengths, we report the data for $H=500$ Oe, which is not
sufficient to saturate the samples, but is large enough to magnetize the
nanoparticles and reveal the Verwey transition and blocking temperature.
After reaching temperatures well above the Verwey transition ($T_V
\sim100-130$ K), but still below the melting point ($\sim270$ K in our
experiments), the system was cooled down without tuning the magnetic field
off (field cooling, FC-C process). Finally, after reaching $T=5$ K, the
magnetic field was turned off and $M\left( T\right) $ was measured again
upon warming. This is annealing of the metastable remanent magnetization
(Rem-W) and it provides important information about the barriers and
anisotropy in the system. It should be noted that we also measured a variety
of other magnetic parameters, including the induced orientation of particles
by a magnetic field when samples were warmed above the melting point and
refrozen in a magnetic field. The results were consistent with the
conclusions of this paper.

Figure \ref{fig1} shows ZFC-W, FC-C and Rem-W measurements in four samples
containing nanoparticles of magnetite and in a high quality synthetic single
magnetite crystal for comparison. The top four frames show data for
biological magnetite nanoparticles in frozen magnetotactic bacterial cells
of strains AMB-1, MC-1, MMS-1 and MV-1, respectively. The images on the
right represent TEM images of the corresponding bacteria and magnetite
particles in magnetosome chains that apparently remained stable after cell
lysis. The lower left frame of Fig.~\ref{fig1} shows data for synthetic
magnetite nanoparticles obtained by slow co-precipitation in agarose gel.
Finally, the lower right image shows similar data measured in a large
synthetic crystal of Fe$_{3}$O$_{4}$. Clearly, the magnetic signature of the
Verwey transition is seen as a sharp change in the magnetic moment at $T_{V}$
in all bacterial samples as well as in the single crystal. Some broadening
of the signal for MC-1 and MMS-1 could still be due to partial oxidation of
magnetite \cite{ozdemir}, but most likely is due to a reduced barrier for
magnetic moment reorientation. Apparently, the difference between FC and ZFC
below $T_{V}$ follows from large magnetic irreversibility below the
transition due to larger magnetocrystalline anisotropy. In addition, larger
remanence can be because of pinning of the magnetic easy axis in a
low-temperature monoclinic phase along the $\langle100\rangle$ direction
closest to the applied field. The transition temperature itself is lower in
magnetite from magnetotactic bacteria when compared to the single crystal,
except for strain MV-1 where it is almost the same. On the other hand,
volume of the magnetite nanocrystals in MV-1 is the smallest (see Table I),
so this implies that size reduction itself does not cause a shift or
broadening of the transition temperature as was reported for synthetic
magnetite nanoparticles \cite{lima,yang}.

One of the best ways to reveal the magnetic signature of the Verwey
transition is to anneal the metastable remanent state by cooling samples
down in a magnetic field and turning it off at low temperature (5 K in our
case). In this case, the thermal energy, $k_{B}T$, only competes with the
dipolar collective barrier and magnetocrystalline anisotropy (no Zeeman term
present). Therefore, a change in anisotropy at $T_{V}$ is well reflected in
the measurements. This is known as (a variation of) the Moskowitz test \cite%
{moskowitz} which has been used to determine whether magnetite particles
from magnetotactic bacteria are present in sediments and water samples.

\begin{figure}[ptb]
\includegraphics[width=9cm]{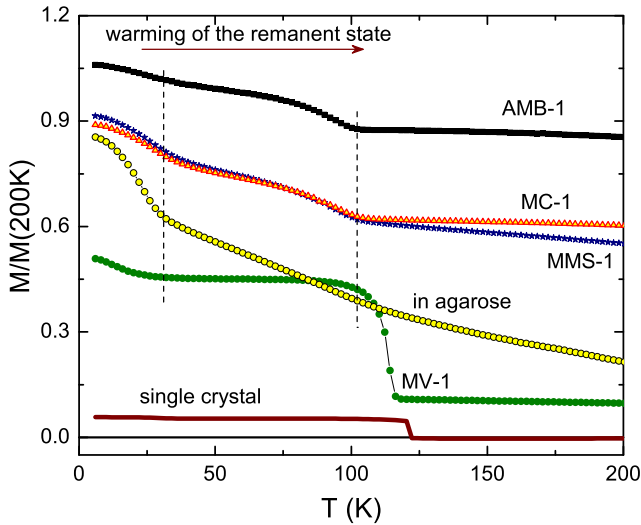}
\caption{Annealing of the remanent state (Rem-W) in six different samples of
magnetite as indicated by labels. Normalization at $200$ K was done using
values obtained from the ZFC-W curves at 500 oe as described in the text.}
\label{fig2}
\end{figure}

Figure \ref{fig2} shows results of such measurements from four bacterial
magnetite samples, synthetic nano-magnetite particles in agarose and in a
large single magnetite crystal. Whereas the single crystal and
nano-magnetite from strain MV-1 show very sharp transitions, the rest of the
bacterial magnetite particles show somewhat broader transitions.
Importantly, the transition is recognizable in the synthetic magnetite
nanoparticles prepared in agarose gel. We interpret these data in terms of
the competition between the magnetic easy axis and shape anisotropy. In
strain MV-1, which has most elongated particles, reorientation and alignment
of the magnetic moments of individual nanocrystals is greatly assisted by
shape anisotropy. In addition, there are two characteristic regions notable
in the plots shown in Fig.~\ref{fig2}. The first is roughly up to $30$ K and
the second is between $30$ K and $T_{V}$ (which varies between $100$ K and $%
125$ K for different samples). The low temperature part, which is absent in
the single crystal, is most likely related to dipolar interparticle
interactions that lead to enhancement of the collective barrier for magnetic
moment reorientation \cite{prozorov99,prozorov04}.

\begin{figure}[ptb]
\begin{center}
\includegraphics[width=9cm]{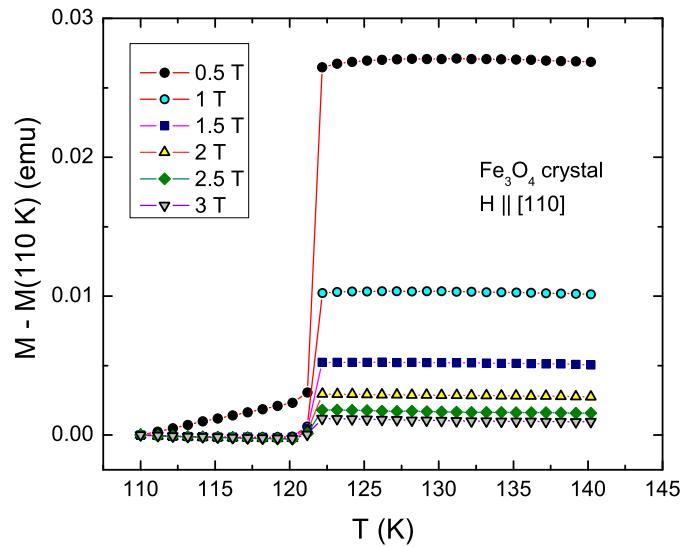}
\end{center}
\caption{Magnetic signature of the Verwey transition in a Fe$_{3}$O$_{4}$
single crystal measured at different magnetic field strengths as indicated
in the graph.}
\label{fig3}
\end{figure}

Another aspect of temperature-dependent magnetization is the absolute value
of the magnetic moment across the Verwey transition. Figure \ref{fig3} shows
a set of transition curves in a single crystal measured in different fields.
The difference between $T<T_{V}$ and $T>T_{V}$ appears to diminish at higher
fields that saturate the magnetic moment. This means that the total magnetic
moment per unit cell remains unchanged and all dramatic changes in
magnetization at lower fields come either from re-orientation of the moments
away from the field axis or a significant change in the magnetic domain
structure.

\subsubsection{Evolution of magnetization in MV-1 magnetite after different
treatments.}

\begin{figure}[tb]
\begin{center}
\includegraphics[width=9cm]{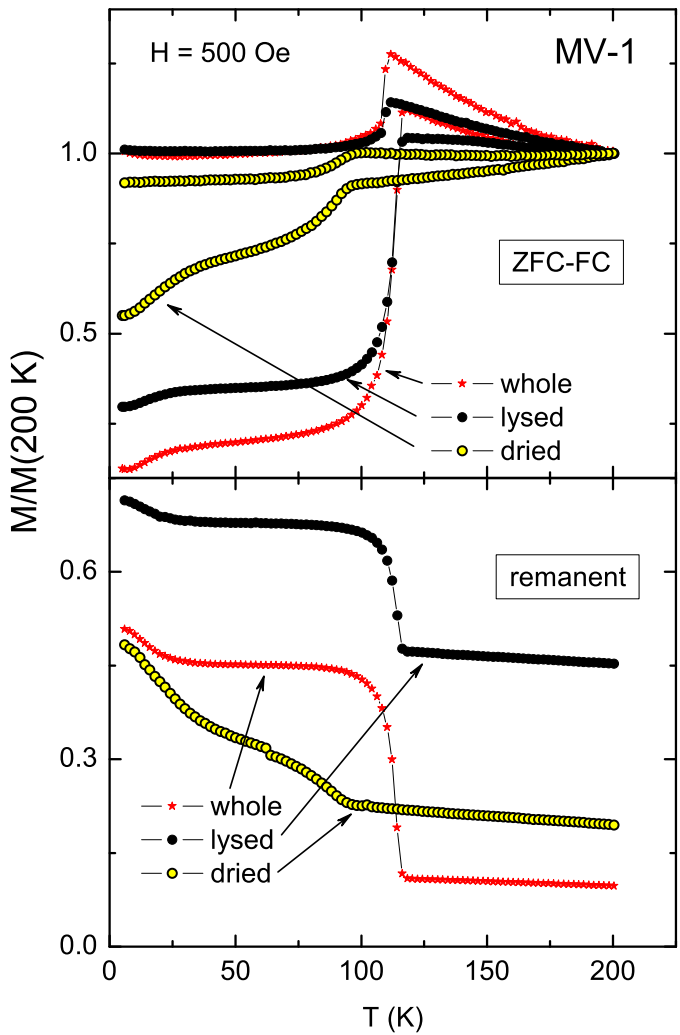}
\end{center}
\caption{$M\left( T\right) $ measurements from magnetite produced by cells
of strain MV-1 with chains (whole and lysed cells) and dried cells with
disrupted chains (individual particles). Top panel shows ZFC-W and FC-C
measurements, whereas bottom panel shows Rem-W data.}
\label{mv1}
\end{figure}

Now we discuss in detail the magnetic measurements obtained in MV-1
magnetite crystals. Magnetite crystals from MV-1 strain have the smallest
volume but are the most elongated of those studied. As evident from Fig.~\ref%
{mv1}, both whole and lysed bacterial cells show very pronounced step-like
feature at $T_{V}$. The top frame of Fig.~\ref{mv1}, shows ZFC-W and FC-C
measurements, whereas the lower frame shows REM-W data. Remanent
magnetization is larger in lysed bacteria due to enhanced interchain
interaction (chains are closer to each other). In contrast to the whole and
lysed nanoparticles, dried powder that contains disassembled chains (but
still preserved randomly-oriented nanocrystals) show significant smearing of
the magnetic signature at the Verwey transition. Importantly, TEM imaging
suggests that individual nanocrystals remain intact and undamaged after the
chain has been disrupted. These results provide the strongest evidence for
the importance of the collective long-range anisotropy in determining the
magnetic response. The chain acts as a single dipole with a very large
anisotropy with the effective blocking temperature much larger than the
Verwey temperature. The anisotropy of individual nanoparticles in a
disassembled chain is not sufficient to prevent thermal randomization and
the magnetic feature at $T_{V}$ is significantly smeared.

\subsection{Magnetization loops}

An important difference between large single magnetite crystals and a
collection of magnetite nanoparticles is revealed in the magnetization loops
measured at low temperatures.

\begin{figure}[ptb]
\begin{center}
\includegraphics[width=9cm]{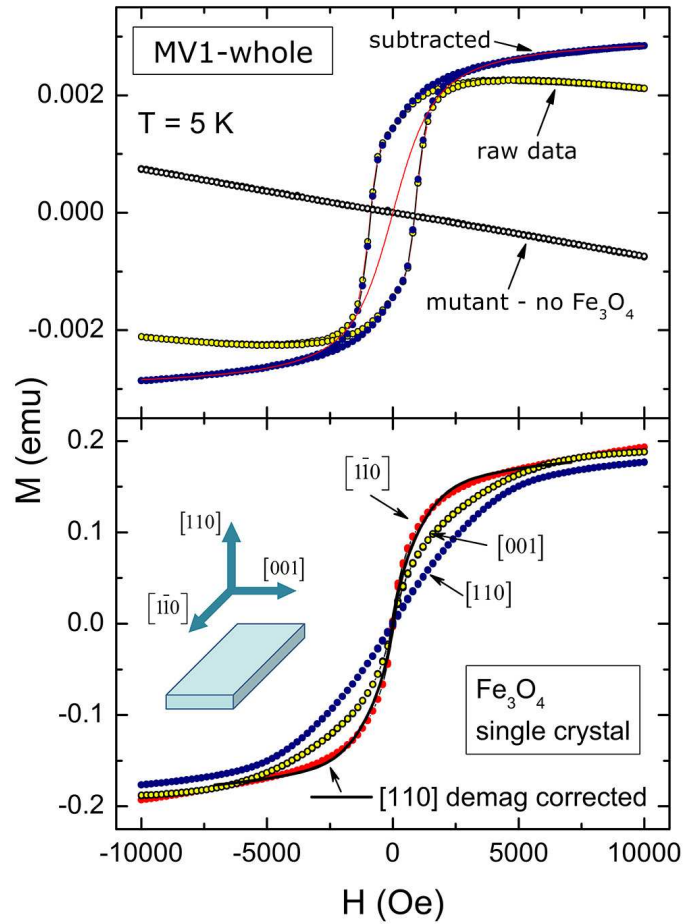}
\end{center}
\caption{$M\left( H\right) $ loops measured at $5$ K in: (\textbf{top})
frozen cell suspension of strain MV-1 whose magnetosomes contain Fe$_3$O$_4$
nanoparticles. The diamagnetic background from cells of a non-magnetic
mutant strain of MV-1 (contains no magnetite) can be subtracted from the
wild-type magnetic strain. The solid curve (red online) is a fit to the
Langevin function. \textbf{(bottom)} single crystal Fe$_{3}$O$_{4}$ and
schematics of its geometry. See discussion in the text.}
\label{fig4}
\end{figure}

The top frame of Fig.~\ref{fig4} shows the $M\left( H\right) $ curve
obtained from a frozen cell suspension of strain MV-1 (see TEM image in Fig.~%
\ref{fig1}). The diamagnetic background comes from the organic and other
material that make up the bacterial cell (everything except the magnetite)
which can be subtracted from measurement of the wild-type (normal) strain by
subtracting the measurement from a non-magnetotactic mutant of MV-1 which
does not produce magnetite \cite{dubbels}. The subtraction results in a
regular magnetization curve the reversible part of which is well-described
by the Langevin function (solid curve, red color online) as expected for
superparamagnetic material. The lower frame shows measurements from the
large single magnetite crystal. The magnetization behaves exactly as
expected for a soft ferromagnet \cite{aharoni} - even simple correction for
demagnetization, $H=H_{applied}-4\pi NM$, works well as shown for the curves
measured in two crystallographically identical orientations, $\left[ 110%
\right] $ and $\left[ 1\overline{1}0\right]$. Schematics of the measured
crystal and its axes is shown in the inset. First, a striking difference
between the single crystals and the nanoparticles is the total absence of
magnetic hysteresis in all orientations. This can be understood in terms of
pinning-free magnetic domains in the single crystal. With the domains,
remagnetization occurs by motion of the domain walls. The domains are absent
in (monodomain) nanoparticles and their remagnetization involves magnetic
moment rotation over a significant energy barrier and this results in
magnetic hysteresis.

\begin{figure}[ptb]
\begin{center}
\includegraphics[width=9cm]{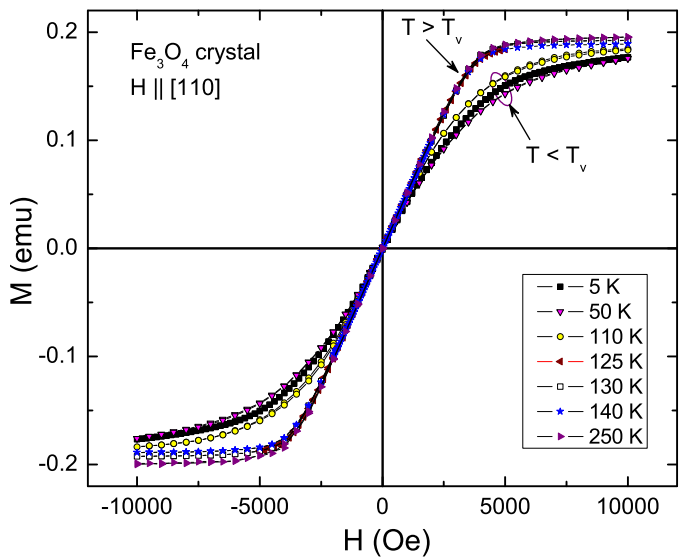}
\end{center}
\caption{Magnetization loops measured at different temperatures in a single
magnetite crystal along the [110] direction. No magnetic hysteresis is
observed and there is a likely as strong change in magnetic anisotropy at
the Verwey transition.}
\label{fig5}
\end{figure}

Another important observation is shown in Fig.~\ref{fig5} which shows three $%
M\left( H\right) $ loops measured in a Fe$_{3}$O$_{4}$ single crystal.
Clearly, there is no detectable magnetic hysteresis at all temperatures.
Furthermore, there is a significant difference in the initial
susceptibility, $dM/dH$, below and above $T_{V}$, but the saturation
magnetization is not at all different. In addition, the initial magnetic
susceptibility is larger at higher temperatures indicating that it is
governed by the temperature-dependent magnetic anisotropy rather than the
Brillouin variable, $\mu H/k_{B}T$, which increases with a decrease in
temperature. This means that the main magnetic effect at the Verwey
transition is due to a significant increase of magnetic anisotropy upon
cooling through $T_{V}$ and not due to a change in the magnetic moment per
unit cell.

\subsubsection{MV-1 - hysteresis loops}

It is possible to gain further insight into the physics of nanophase
magnetite by comparing $M\left( H\right) $ measurements performed on the
same strain of magnetotactic bacteria treated differently. For example,
whole, intact bacterial cells can be directly compared to lysed cells whose
cell membranes are destroyed and only magnetosome chains containing
individual Fe$_{3}$O$_{4}$ nanocrystals remain. (Magnetosome membranes are
still present even when cells are lysed and this keeps chains intact).
Depending on how cells are lysed, chains may be disrupted and become
shorter. The next step is to eliminate all organic material which would
leave only individual magnetite nanocrystals.

\begin{figure}[ptb]
\begin{center}
\includegraphics[width=9cm]{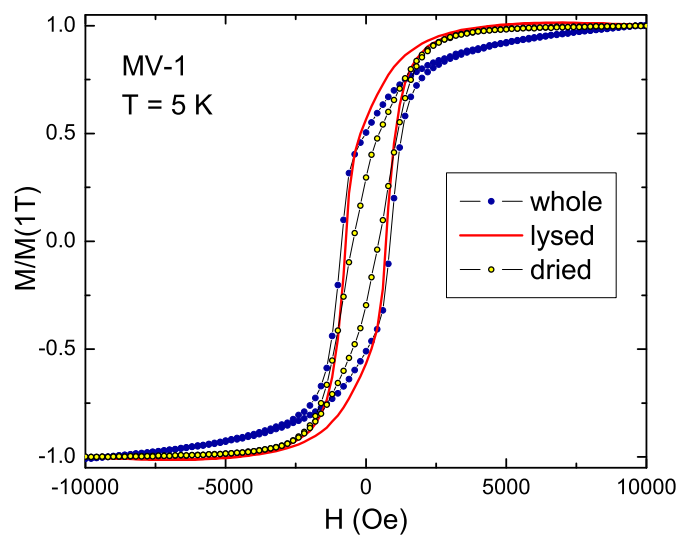}
\end{center}
\caption{Comparison of the $5$ K $M\left( H\right) $ hysteresis loops from
the whole, lysed and freeze-dried cells from strain MV-1 which contain
magnetite nanoparticles. }
\label{fig6}
\end{figure}

Figure \ref{fig6} shows the result of this comparison. Clearly, there is a
reduction of hysteresis in the freeze-dried cells. The saturation field,
however, is similar in lysed and whole samples. This leads to the important
conclusion that a significant part of the magnetic hysteresis is caused by
long-range (dipolar) interchain interactions and the shape anisotropy
contributions of the entire long chains and not by particle agglomeration or
random dipolar interactions.

\subsection{Magnetic relaxation}

Significant magnetic relaxation is usually observed in assemblies of
magnetic nanoparticles below the blocking temperature \cite%
{prozorov99,prozorov04}. It is believed that this relaxation is a result of
Arrhenius thermal activation, $\exp\left( -U/k_{B}T\right)$ where $U$ is the
barrier for magnetic moment re-orientation. In a simple single-particle
barrier, $U$ depends only on magnetocrystalline anisotropy and the strength
of the applied field, so the relaxation should be time-exponential. However,
in assemblies of nanoparticles, relaxation is always time-logarithmic, which
we show is due to the collective nature of the barrier that now depends on
the total magnetic moment, $U\left( M\right)$ \cite{prozorov99,prozorov04}.
Alternative theories invoke size (hence barrier) distributions that
supposedly lead to stretched exponential relaxation. Magnetotactic bacterial
nano-crystalline magnetite provides useful insight into the problem because
their size distribution is very narrow and their crystal structure is
perfect (that excludes barrier variations due to the amorphous nature of the
synthetic nanoparticles).

\begin{figure}[ptb]
\begin{center}
\includegraphics[width=9cm]{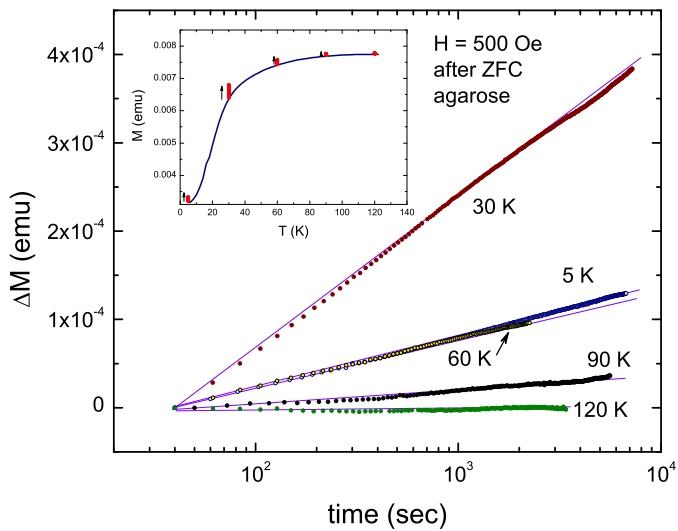}
\end{center}
\caption{Magnetic relaxation in synthetic magnetite nanoparticles (in
agarose) measured at $H=500$ Oe applied after ZFC to the indicated
temperature. Inset shows the corresponding ZFC-W temperature scan.}
\label{fig7}
\end{figure}

Figure \ref{fig7} shows magnetic relaxation in synthetic magnetite
nanoparticles in agarose gel. Each curve is obtained by measuring $M$ versus
time after a magnetic field of $500$ Oe was applied after cooling in a zero
field to $5,30,60,90$ and $120$ K. The inset shows the same relaxation
curves plotted in $M-T$ coordinates. Arrows indicate the direction of the
increase in time. Very similar relaxation curves are observed in frozen
ferrofluids \cite{prozorov04} and dry powders \cite{prozorov99}. Our
previous work has shown, both experimentally and theoretically, that obvious
time-logarithmic dependence of magnetization is due to the collective
barrier for magnetic relaxation. Single-particle barrier does not depend on
the total magnetic moment (of the surroundings) and results in
time-exponential relaxation. With dipolar interactions in the system, the
barrier is determined within correlation volume that depends explicitly on
the total magnetic moment. It can be quite generally shown that this is
sufficient to produce time-logarithmic magnetic relaxation \cite%
{prozorov99,prozorov04}.

\subsubsection{MV-1 - magnetic relaxation}

If our interpretation of the $M\left( T\right) $ and $M\left( H\right)$
measurements (that there is a very large magnetic anisotropy in the
magnetite nanoparticle chains) is correct, we should not observe significant
magnetic dynamics. However, when chains are disrupted, we should observe
significant magnetic relaxation similar to that shown in Fig.~\ref{fig7}.

\begin{figure}[ptb]
\begin{center}
\includegraphics[width=9cm]{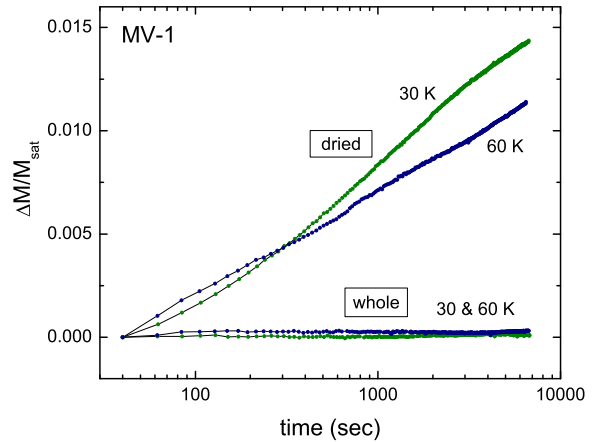}
\end{center}
\caption{Magnetic relaxation curves measured in magnetite nanocrystals
produced by strain MV-1 at $H=500$ Oe applied after ZFC. The two lower lines
are for intact frozen bacterial cells, whereas the upper lines are for dried
cells with disassembled chains.}
\label{fig8}
\end{figure}

Figure \ref{fig8} shows magnetic relaxation curves measured in magnetite
nanocrystals produced by strain MV-1. The two lower lines are for whole
frozen bacterial cells, whereas the upper lines are for freeze-dried
bacteria cells where the chains are disrupted. Particle morphology was
preserved as determined by TEM imaging. The curves were normalized by the
saturation magnetization (in a $5$ Tesla field) at $5$ K. There is a clear
difference between magnetic relaxation trends in these cases. It should be
noted that we attempted to measure magnetic relaxation in all our specimens
with the same result - no noticeable magnetic relaxation is observed in
magnetic nanoparticles organized in magnetosome chains (in whole and lysed
cells) and significant relaxation is seen in dried cells where chains were
disrupted. This suggests that single-particle barrier or, as previously
believed, a distribution of sizes (that would lead to so-called stretched
exponential behavior), cannot explain the observed time logarithmic
relaxation. Instead, random long-range dipolar interparticle interactions
that lead to magnetic moment - dependent collective barrier explain the
phenomena \cite{prozorov99,prozorov04}.

\section{Discussion}

Our results indicate that,

\begin{enumerate}
\item A sharp magnetic signature at the Verwey transition is clearly present
in magnetite nanocrystals organized in  chains. However, it is significantly
smeared in separated nanocrystals.

\item Measurements of magnetic relaxation suggest that chains of
magnetosomes behave as very large magnetic dipoles  with the intra-chain
magnetic induction aligned along their axis that corresponds to the $\left[%
111\right] $  direction.

\item Randomly-arranged nanoparticles (not in chains) show a pronounced
time-logarithmic magnetic relaxation as  expected from the collective nature
of the barrier for reorientation of the magnetic moment.

\item Although the degree of crystallinity in magnetite nanoparticles
certainly plays a role (no magnetic signature  at $T_{V}$ was found in
ferritin-templated amorphous nanoparticles \cite{tprozorov}), the Verwey
transition  remains smeared even in perfect, but uncorrelated nanoparticles.

\item The magnetic signature of the Verwey transition does not correlate
with the mean particle size (or volume) at  least in the 30-50 nm size
range. Instead, it seems to depend greatly on particle shape and chain
anisotropy.
\end{enumerate}

We explain these observations as follows. At the Verwey transition, the
magnetic moment does not change in magnitude, but changes direction due to
the switching of the magnetic easy axis. In bulk crystals,
magnetocrystalline anisotropy is still much larger than thermal activation
energy, $k_{B}T$, so when switching of the magnetic moment occurs, it
appears as an abrupt change in the magnetization (always measured along some
chosen axis, so it always reflects the rotation). In contrast, in individual
nanoparticles, magnetic fluctuations are strong enough to overcome the Ne%
\'{e}l barrier, $KV$ (where $K$ is magnetic anisotropy and $V$ is particle
volume), and randomize the magnetic moment (spatially and temporally). In
this case, reorientation cannot be considered because the magnetic moment is
continually fluctuating. As a result, only a very small and diffuse change
is observed due to the finite distribution of particle barriers and
directions. Nanoparticles organized in chains act as long dipoles with
enhanced effective anisotropy along the chain and thermal fluctuations are
insufficient to overcome this barrier. Phenomenologically, it is to say that
the Verwey temperature is larger than the blocking temperature of individual
nanoparticles, $T_{V}>T_{B}$, but less than the effective blocking
temperature of the chains, $T_{V}<T_{B}^{chains}$. This is an important
conclusion, because it implies that the mechanism of the Verwey transition
does not depend on any particular orientation of the internal magnetic field
with respect to the crystal structure (in addition, the $\left[ 111 \right]$
direction is degenerate with respect to any of the principal cubic axes
along which charge ordering may occur). On the other hand, it also implies
that the Verwey transition occurs in nanoparticles at temperatures close to
bulk values that rules out proposed long-range magneto-elastic coupling
leading to the disappearance of monoclinic distortion and the Verwey
transition altogether.

\begin{acknowledgments}
We thank B. M. Moskowitz, R. B. Frankel and R. J. McQueeney for useful
discussions and suggestions. We thank J. Honig for careful reading of the
manuscript, discussions and providing high-quality Fe$_{3}$O$_{4}$ single
crystals. R.P. thanks his cat Physya for support. Work at the Ames
Laboratory was supported by the Department of Energy-Basic Energy Sciences
under Contract No. DE-AC02-07CH11358. R. P. acknowledges support from the
NSF grant number DMR-06-03841 and the Alfred P. Sloan Foundation. D.A.B. and
T.J.W. acknowledge support from the NSF grant number EAR-0311950.
\end{acknowledgments}


\begin{thebibliography}{99}
\bibitem{parks} G. S. Parks and K. K. Kelley, J. Phys. Chem. \textbf{30}, 47
(1926).

\bibitem{verwey39} E. J. W. Verwey, Nature (London, United Kingdom) \textbf{%
144}, 327 (1939).

\bibitem{review} (special volume on the Verwey transition) Philos. Mag. B 
\textbf{42}, 325 (1980).

\bibitem{muxworthy} A. R. Muxworthy and E. McClelland, Geophysical Journal
International \textbf{140}, 101 (2000).

\bibitem{garcia} J. Garc\'{\i}a and G. Sub\'{\i}as, Journal of Physics:
Condensed Matter \textbf{16}, R145 (2004).

\bibitem{leonov} I. Leonov and A. N. Yaresko, Journal of Physics: Condensed
Matter \textbf{19}, 021001 (2007).

\bibitem{verwey41} E. J. Verwey, and P. W. Haayman, Physica \textbf{9}, 979
(1941).

\bibitem{lima} E. Lima, Jr., A. L. Brandl, A. D. Arelaro, and G. F. Goya, J.
Appl. Phys. \textbf{99}, 083908 (2006).

\bibitem{wang90} P. Wang, M. W. Wittenauer, D. J. Buttrey, R. W. Choi, P.
Metcalf, Z. Kakol and J. M. Honig, J. Crystal  Growth \textbf{104}, 285
(1990).

\bibitem{kakol} Z. Kakol and J. M. Honig, Phys. Rev. B \textbf{40}, 9090
(1989).

\bibitem{wang} J. Wang, Q. Chen, X. Li, L. Shi, Z. Peng, and C. Zeng,
Chemical Physics Letters 390, 55 (2004).

\bibitem{goya} G. F. Goya, T. S. Berquo, F. C. Fonseca, and M. P. Morales,
Journal of Applied Physics \textbf{94}, 3520  (2003).

\bibitem{arelaro} A. D. Arelaro, A. L. Brandl, E. Lima, Jr., L. F. Gamarra,
G. E. S. Brito, W. M. Pontuschka, and G. F.  Goya, Journal of Applied
Physics \textbf{97}, 10J316 (2005).

\bibitem{markovich} G. Markovich, T. Fried, P. Poddar, A. Sharoni, D. Katz,
T. Wizansky, and O. Millo, MRS Symposium  Proceedings \textbf{746}, 151
(2003).

\bibitem{blakemore} R. P. Blakemore, Science \textbf{190}, 377 (1975).

\bibitem{frankel75} R. B. Frankel, R. P. Blakemore, R. S. Wolfe, Science 
\textbf{203}, 1355 (1979).

\bibitem{krueger} S. Krueger, G. J. Olson, J. J. Rhyne, R. P. Blakemore, Y.
A. Gorby, and N. Blakemore, J. Mag. Mag.  Mater. \textbf{82}, 17 (1989).

\bibitem{moskowitz} B. M. Moskowitz, R. B. Frankel, D. A. Bazylinski, H. W.
Jannasch, and D. R. Lovley, Geophys. Res.  Lett. \textbf{16}, 665 (1989).

\bibitem{meldrum} F. C. Meldrum, S. Mann, B. R. Heywood, R. B. Frankel, and
D. A. Bazylinski, Proceedings: Biological  Sciences \textbf{251}, 237 (1993).

\bibitem{penninga} I. Penninga, H. de Waard, B. M. Moskowitz, D. A.
Bazylinski, and R. B. Frankel, J. Mag. Mag. Mater.  \textbf{149}, 279 (1995).

\bibitem{bazylinski} D. A. Bazylinski, and R. B. Frankel, Nature Reviews
Microbiology \textbf{2}, 217 (2004).

\bibitem{frankel} R. B. Frankel, D. A. Bazylinski, M. Johnson, and B. L.
Taylor, Biophys. J. \textbf{73}, 994 (1997).

\bibitem{wolin} E. A. Wolin, M. J. Wolin and R. S. Wolfe, J. Biol. Chem. 
\textbf{238}, 2882 (1963).

\bibitem{weiss} B. P. Weiss, S. S. Kim, J. L. Kirschvink, R. E. Kopp, M.
Sankaran, A. Kobayashi, and A. Komeili, Earth  and Planetary Science Letters 
\textbf{224}, 73 (2004).

\bibitem{faivre} D. Faivre, N. Menguy, F. Guyot, O. Lopez, and P. Zuddas,
American Mineralogist \textbf{90}, 1793  (2005).

\bibitem{pan} Y. Pan, N. Petersen, M. Winklhofer, A. F. Davila, Q. Liu, T.
Frederichs, M. Hanzlik, and R. Zhu, Earth  and Planetary Science Letters 
\textbf{237}, 311 (2005).

\bibitem{polyakova} T. Polyakova, and V. Zablotskii, Journal of Magnetism
and Magnetic Materials 293, 365 (2005).

\bibitem{simpson} E. T. Simpson, T. Kasama, M. Posfai, P. R. Buseck, R. J.
Harrison, and R. E. Dunin-Borkowski, Journal  of Physics: Conference Series 
\textbf{17}, 108 (2005).

\bibitem{ozdemir} \"{O}. \"{O}zdemir, D.J. Dunlop, B. M. Moskowitz, Geophys.
Res. Lett. \textbf{20, }1671 (1993).

\bibitem{kobayashi} A. Kobayashi, J. L. Kirschvink, C. Z. Nash, R. E. Kopp,
D. A. Sauer, L. E. Bertani, W. F. Voorhout,  and T. Taguchi, Earth and
Planetary Science Letters \textbf{245}, 538 (2006).

\bibitem{weyland} M. Weyland, T. J. V. Yates, R. E. Dunin-Borkowski, L.
Laffont, and P. A. Midgley, Scripta Materialia  \textbf{55}, 29 (2006).

\bibitem{bazylinski07} D. A. Bazylinski, and T. J. Williams in \textit{%
"Magnetoreception and Magnetosomes in Bacteria"}, edited by D. Sch\"{u}ler
(Springer-Verlag, Berlin, Heidelberg, 2007), pp. 37-75.

\bibitem{bazylinski88} D. A. Bazylinski, R. B. Frankel, and H. W. Jannasch,
Nature \textbf{334}, 518 (1988).

\bibitem{williams06} T. J. Williams, C. L. Zhang, J. H. Scott, and D. A.
Bazylinski, Appl. Environ. Microbiol.  \textbf{72}, 1322 (2006).

\bibitem{matsunaga} T. Matsunaga, T. Sakaguchi, and F. Tadokoro, Appl.
Microbiol. Biotechnol. \textbf{35}, 651 (1991).

\bibitem{dubbels} B. L. Dubbels, A. A. DiSpirito, J. D. Morton, J. D.
Semrau, J. N. Neto, and D. A. Bazylinski,  Microbiology \textbf{150}, 2931
(2004).

\bibitem{blakemore79} R. P. Blakemore, D. Maratea and R. S. Wolfe, J
Bacteriol \textbf{140}, 720 (1979).

\bibitem{prozorov99} R. Prozorov, Y. Yeshurun, T. Prozorov, and A. Gedanken,
Phys. Rev. B \textbf{59}, 6956 (1999).

\bibitem{prozorov04} R. Prozorov and T. Prozorov, J. Mag. Mag. Mater. 281, 
\textbf{312} (2004).

\bibitem{aharoni} A. Aharoni, \textit{"Introduction to the Theory of
Ferromagnetism"} (Oxford University Press, 2001).

\bibitem{snezhko05} A. Snezhko, T. Prozorov, and R. Prozorov, Phys. Rev. B 
\textbf{71}, 024527 (2005).

\bibitem{cao97} X. Cao, R. Prozorov, Y. Koltypin, G. Kataby, I. Felner, and
A. Gedanken, J. Mater. Res. \textbf{12},  402 (1997).

\bibitem{cao97a} X. Cao, Y. Koltypin, R. Prozorov, G. Kataby, and A.
Gedanken, J. Mater. Chem. \textbf{7}, 2447 (1997).

\bibitem{cao97b} X. Cao, Y. Koltypin, G. Katabi, R. Prozorov, I. Felner, and
A. Gedanken, J. Mater. Chem. \textbf{7},  1007 (1997).

\bibitem{yang} J. B. Yang, X. D. Zhou, W. B. Yelon, W. J. James, Q. Cai, K.
V. Gopalakrishnan, S. K. Malik, X. C. Sun,  and D. E. Nikles, J. Appl. Phys. 
\textbf{95}, 7540 (2004).

\bibitem{tprozorov} T. Prozorov, S. K. Mallapragada, B. Narasimhan, L. Wang,
P. Palo, M. Nilsen-Hamilton, T. J.  Williams, D. A. Bazylinski, R. Prozorov
and P. C. Canfield, Adv. Funct. Mater. \textbf{17}, 951 (2007).
\end{thebibliography}
\end{document}